\documentclass[aps, prl,reprint,a4paper,superscriptaddress,amsmath,amsfonts,amssymb, 
floatfix, 
]{revtex4-2}
\usepackage{bm}
\usepackage[caption=false]{subfig}
\usepackage{tikz}
\usetikzlibrary{shapes.misc,positioning,shapes}
\usetikzlibrary{patterns}
\usepackage{array, multirow}
\usepackage[colorlinks,linkcolor=blue,citecolor=blue]{hyperref}%
\usepackage{relsize}

\usepackage{times}

\usepackage[normalem]{ulem}
\usepackage[english]{babel}
\usepackage{blindtext}
\begin{document}
\title{Precision Quantum Parameter Inference with Continuous Observation}

\author{Bijita Sarma}
\email{bijita.sarma@fau.de}
\affiliation{Friedrich-Alexander-Universit\"at Erlangen-N\"urnberg, Staudtstra{\ss}e 7, 91058 Erlangen, Germany}

\author{Junxin Chen}
\affiliation{\textit{LIGO}, Massachusetts Institute of Technology, Cambridge, MA 02139, USA}

\author{Sangkha Borah}
\email{sangkha.borah@mpl.mpg.de}
\affiliation{Max Planck Institute for the Science of Light, Staudtstra{\ss}e 2, 91058 Erlangen, Germany}
\affiliation{Friedrich-Alexander-Universit\"at Erlangen-N\"urnberg, Staudtstra{\ss}e 7, 91058 Erlangen, Germany}

% TC:ignore

\begin{abstract}
Quantum Parameter Estimation (QPE) is important from the perspective of both fundamental quantum research and various practical applications of quantum technologies such as for developing optimal quantum control strategies. Standard and traditional methods for QPE involve projective measurements on thousands of identically prepared quantum systems. However, these methods face limitations, particularly in terms of the required number of samples and the associated experimental resources. In this work, we present a novel method for precise QPE that diverges from conventional techniques, employs continuous measurements, and enables accurate QPE with a single quantum trajectory. In an application, we demonstrate the use of the method for the task of parameter estimation and force sensing of a levitated nanoparticle. 
\end{abstract}
\maketitle
%TC:endignore

An experiment of a quantum system typically involves three major steps: (a) preparation and initialization, (b) measurement, and (c) analysis. During the preparation and initialization step, the measurement system is isolated and prepared (in a given state) and the measurement apparatus is initialized (reset)~\cite{WisemanMiburnBook,KurtJacobsBook}. Measurement involves coupling the system to the apparatus and recording the measurement results. In the analysis step, the measurement records are processed and analyzed to extract relevant information about the quantum system under study. This process can involve statistical analysis, comparison with theoretical predictions, and error estimation~\cite{Giovannetti2011, Giovannetti2004, Zhang2022Oct}. The preparation procedure can be described by a set of classical parameters or settings of a physical device, and the measurement results encode information correlated with the preparation and initialization procedure. Often, these parameters are not known in advance or are only partially known, and the task of quantum parameter estimation (QPE) is simply to reliably estimate them as accurately as possible, often with post-processing of the measured data in the analysis phase\cite{Barbieri2022Jan_PRX_Metrology_Review, Degen2017, Braun2018}. The term ``quantum" in QPE comes from the fact that it is a quantum system that facilitates the transfer of information from the classical parameters of the preparation process to the classical measurement results~\cite{WisemanMiburnBook, Sidhu2020, Pezze2018}. \\
\indent As quantum technology continues to advance, QPE is expected to play a crucial role in harnessing the power of quantum systems for practical applications. In the context of quantum computing, accurate estimation of system parameters can help optimizing the accuracy of various control protocols, including quantum error correction (QEC) and error mitigation, as well as for optimizing the performance and efficiency of various quantum algorithms~\cite{Harper2020Dec, Eisert2020Jul, Meyer2021Jun, Takagi2022Sep,nielsen2001quantum}. Accurate QPE enables a precise understanding of the model of a quantum system, which can be used to develop accurate QEC techniques and other quantum control methods~\cite{Puviani2023GKP, Borah2022QEC}. This becomes particularly relevant in light of recent advances in applying machine learning (ML) to quantum physics problems, in which model-based ML techniques can be utilized to discover optimal control and feedback strategies, despite the presence of measurement noise~\cite{Gebhart2023Mar, Puviani2023GKP}. Additionally, QPE plays a crucial role in quantum sensing and metrology, where precise parameter estimation is necessary for high-precision measurements and the detection of quantum signals~\cite{WisemanMiburnBook, Sidhu2020, Pezze2018, Giovannetti2011, Giovannetti2004, Zhang2022Oct}, which empowers both applications and test of fundamental physics. 
\\
\indent While the preparation procedure may be well understood, the measurement results often exhibit statistical behavior because of the fundamental principles of quantum mechanics. As a result, a single preparation and measurement step is often insufficient to accurately estimate the parameters~\cite{Braun2018Sep}. To address this, it is common practice to repeat the preparation and measurement process in multiple identically prepared systems to accumulate statistical data and make more reliable estimations of the parameters of interest. However, this approach can be time-consuming and resource-intensive and introduces the potential for errors and inconsistencies while preparing and initializing the ensemble of quantum systems and measurement devices. Therefore, researchers are constantly seeking new techniques and technologies to improve the efficiency and accuracy of quantum measurements~\cite{Sidhu2020, Pezze2018, Zhang2022Oct, Ansel2024Jan, Barbieri2022Jan_PRX_Metrology_Review, Degen2017, Braun2018}. One such technique is QPE with continuous measurement~\cite{WisemanMiburnBook, Zhang2017Mar, KurtJacobsBook}. While most previous studies have focused on QPE with projective measurement, there have been some studies that employed continuous measurement of quantum systems~\cite{Rossi2020Nov, Fallani2022Apr, Kiilerich2016Sep, Albarelli2017Dec}. Continuous measurement has the advantage of not fully collapsing a quantum state, but it is limited by measurement noise and measurement-induced backaction. Measurement noise makes it difficult to accurately separate the signal from noise, so continuous measurement QPE methods rely on sampling hundreds of trajectories for estimation. In this work, we propose a method that can accurately determine the precise parameter from a single trajectory in an experiment. We also demonstrate the usefulness of this approach in parameter estimation and force sensing of a levitated nanoparticle.\\
\begin{figure}[t]
    \centering
    \includegraphics[width=1.0\linewidth]{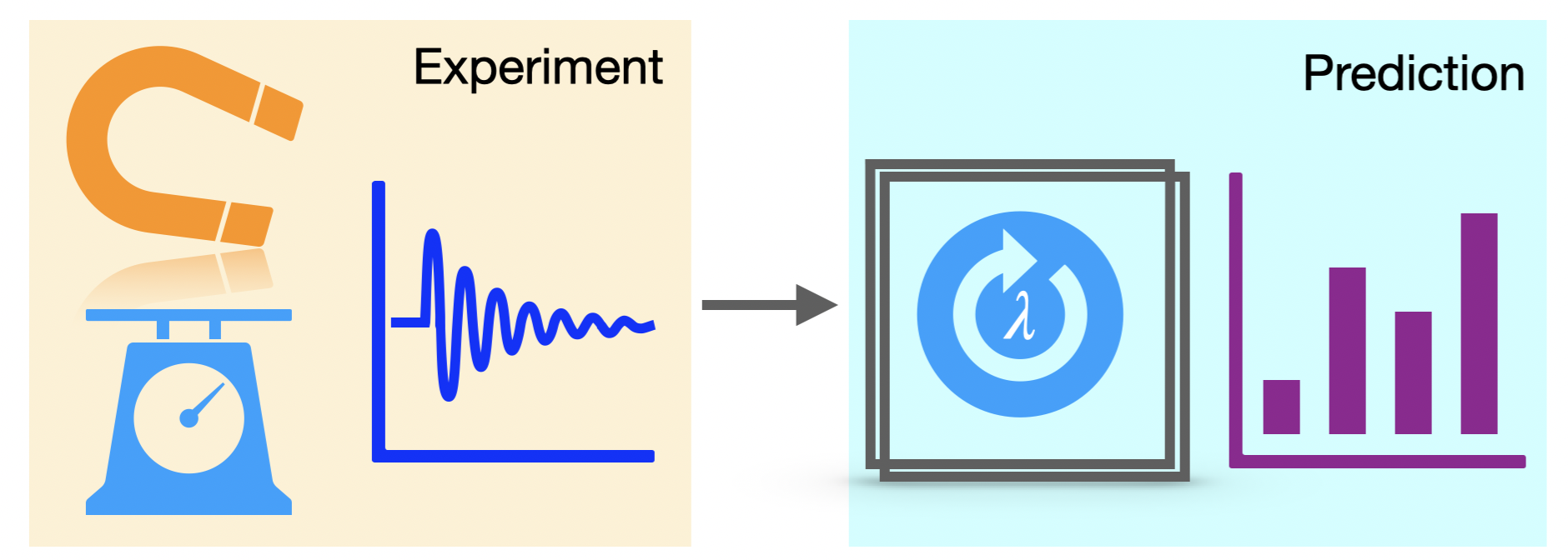}
    \caption{Schematic of the proposed scheme for QPE of a unknown parameter $\lambda_{0}$, encoded in the Hamiltonian of the system, $H(\lambda_0)$. Continuous measurement of the system is performed on the experimental setup (left) to collect a single trajectory of duration $\tau$ with a total of $N$ noisy measurement outcomes. The parameter to be predicted, $\lambda_{0}$ is obtained by simulating offline trajectory with varying $\lambda$. The protocol guarantees that for $\lambda = \lambda_{0}$, the experimental trajectory is perfectly matched by the simulated trajectory, leading to the loss, $|\Delta\mathcal{M}|^2$ [Eq.~\ref{eq:loss}] to be zero (minimum). The same model can also be used to predict other system and detector properties that are not directly encoded into the Hamiltonian, such as the unknown decoherence rates, measurement efficiency, etc.}
    \label{fig:1}
\end{figure}
The schematic of the QPE scheme is shown in Fig.~\ref{fig:1}.  In this approach, a trajectory is collected via continuous measurement in an experiment (left) and given to post-processing (right). The trajectory consists of time series dataset given by ${D} = \{(t_0, \mathcal I_0), ..., (t_j, \mathcal I_j), ...(t_{N-1},\mathcal I_{N-1} )\}$, where $\mathcal I_j$  denotes the noisy measurement data of the measurement operator $\mathcal A$ at time $t_j$.  The post-processing task is to estimate the unknown system parameter $\lambda$ at its optimally accurate value $\lambda_{\rm optimal} \sim \lambda_{0}$, where $\lambda_0$ is the accurate value of the parameter. With the procedure discussed in the following, $\lambda_{\rm optimal}$ can be accurately determined by running a simulation with the quantum master equation (QME) by varying $\lambda$ values and comparing them with the reference trajectory obtained from the experiment. The minimum error is to perfect zero for $\Delta \lambda = 0$, where $\Delta \lambda = |\lambda - \lambda_0|$. The detail of the method is discussed below, with a brief introduction to the continuous measurement process. \\
\indent Continuous quantum measurement involves weak monitoring of a quantum system and is essentially achieved through weak coupling of the measurement device with the system. Suppose that the laboratory quantum system, with Hamiltonian $\hat{H}$, is continuously measured with a weak probe for the measurement operator $\hat{\mathcal A}$ (suitably scaled to make it dimensionless). Such a continuous measurement process leads to conditional stochastic dynamics of the system density matrix in time $\rho_c(t)$ and is described by the so-called quantum stochastic master equation (SME),
\begin{align}
\nonumber
{d\rho_c(t)} =  - i [\hat{H}, ~\rho_c(t)] dt + \kappa \mathcal{D}[\hat{A}] \rho_c(t) dt
\\+ \sqrt{\kappa \eta} \mathcal{H}[\hat{A}] \rho_c(t) d\mathcal{W}(t).
\label{eq:SME}
\end{align}
Here, $\kappa$ is the measurement rate (the rate at which information is extracted from the detector), $\eta$ is the measurement efficiency of the detector and $d\mathcal{W}(t)$ represents an instantaneous random Wiener noise increment (white noise model with zero mean and variance ${dt}$). $\mathcal{D}[\hat{A}]$ and $\mathcal{H}[\hat{A}]$ are the superoperators that describe, respectively, the backaction and diffusion terms in the SME~\cite{WisemanMiburnBook}.  Probing the system with a weakly coupled meter that, in effect, has a broad probability distribution of the quantum state leads to noisy measurement as a time series dataset ${D} = \{ (t, \mathcal{I}_{\rm expt}(t))\}$, which is used for the post-processing step given below.   \\
\indent The QPE estimator set up a simulation with unknown parameters with the modified SME
\begin{align}
\nonumber
d\rho_c(t) = & - i [\hat H(\lambda), ~\rho_c(t)] dt + \kappa \mathcal{D}[\hat{A}] \rho_c(t) dt \\
+& {2\kappa \eta}\left[
{\cal I}_{\rm expt}(t) - \langle \hat A(t)\rangle_c  \right] \mathcal{H}[\hat{A}] \rho_c(t) dt,
\label{eq:mbe_scheme_main_eq}
\end{align}
where $H(\lambda)$ is the Hamiltonian of the system in terms of the unknown parameter $\lambda$. $\mathcal{I}_{\rm expt}(t)$ is the noisy measurement current from $\mathcal{I}_{\rm expt}$ the trajectory collected from the experiment, and $\langle \hat{A}(t) \rangle_c = \mathrm{Tr}[\rho_c \hat{A}]$ is the conditional mean calculated locally at time $t$ during simulation.  
Given an estimator modeled according to Eq.~\ref{eq:mbe_scheme_main_eq} and initialized with a random quantum state $\rho(0)$, it is guaranteed that it converges to the true quantum state $\rho_{\rm expt}(t)$ after a sufficiently long time $t$ [of the order of $t > \kappa^{-1}$]~\cite{WisemanMiburnBook,Rossi2019Oct,Borah2023_no_collapse,Zhang2017Mar}. The time of convergence $t$ is reduced as a function of the increasing fidelity between the initial states of the estimator $\rho(0)$ and the real system $\rho_{\rm expt}(0)$, i.e. $\mathcal{F}(\rho_{\rm expt}(0), \rho(0))$. In fact, the estimator would immediately start following the correct dynamics of the real system if $\mathcal{F}(\rho_{\rm expt}(0), \rho(0))=1$. This fact can be utilized to form a powerful and useful QPE protocol.  
Considering the task of estimating an unknown parameter $\lambda$ encoded in the Hamiltonian $\hat H = \hat H(\lambda)$, the method would include the use of this simulation with varying $\lambda$ [Eq.~\ref{eq:mbe_scheme_main_eq}], providing a straightforward but innovative approach to the QPE of the parameter $\lambda$. This approach can likewise be applied to calculate other pertinent systems and detector parameters, such as the decoherence rate and measurement efficiency, even though these parameters are not directly integrated into the system's Hamiltonian (see discussions below).\\
\indent Essentially, the task at hand is to calculate the following loss function 
\begin{equation}
{\Delta \mathcal{M}^2} = \sum_{t} |\mathcal{I}_{\rm expt}(t) - \mathcal{I}_{\rm QPE}(t)|^2,
\label{eq:loss}
\end{equation}
where $\mathcal I_{\rm QPE}(t)$ is the measurement current  at timestamp $t$ as obtained by the simulator and $\mathcal I_{\rm expt}(t)$ is the same collected from the experiment earlier.  

\begin{figure}[t]
    \centering
    \includegraphics[width=1.0\linewidth]{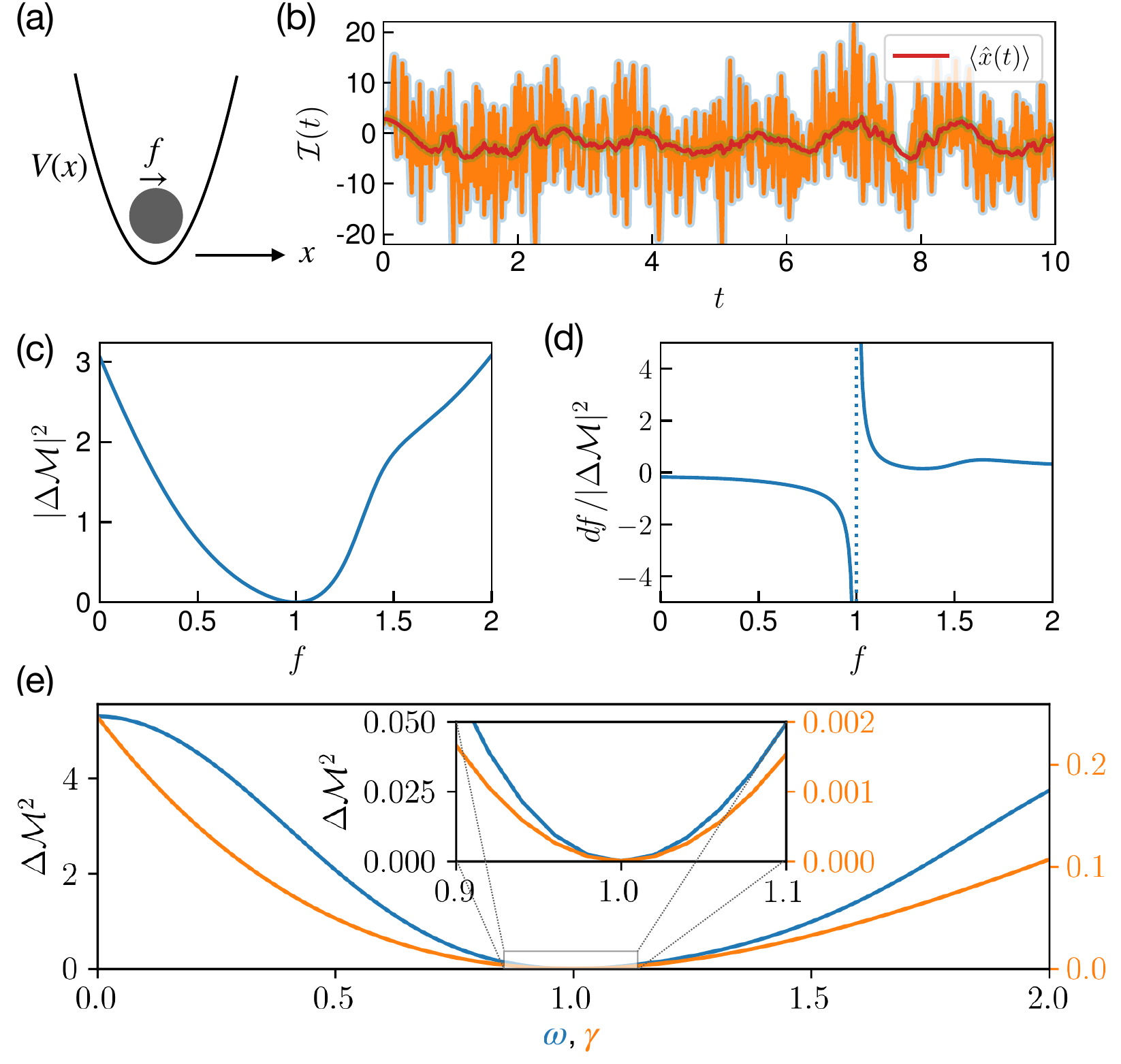}
    \caption{The proposed QPE scheme is applied for force sensing of a levitated nanoparticle. (a) A nanoparticle of mass $m$ is trapped in a potential $V(x)$. An unknown force $\vec {F} = \hbar\omega_0 f\hat{x}$ is applied to it. The task is to estimate the unknown force component $f$. Here, we have re-scaled $f$ so that the correct value of the force is $f_0=1$. (b) A single trajectory $\mathcal{I}_{\rm expt}(t)$ generated by an experiment with the setup of the levitated particle. (c) The loss [Eq. \ref{eq:loss}] calculated as a post-processing step by modeling the system on a computer and using the experimental trajectory as a target. The minimum of the loss should ideally be zero, which makes each point of the experimental trajectory [in (b)] reproduced, yielding ideally zero loss. In effect, the conditional means $\langle \hat{x}(t)\rangle$ also perfectly overlap as shown in (b). (d) The derivative $df/|\Delta \mathcal{M}|^2$ with respect to $f$ is plotted. The sharp peak at $f=1$ shows that the estimation is rather very accurate, which can be enhanced by choosing finer grids for $f$. (e) QPE for determining the frequency ($\omega$) and decoherence rate ($\gamma$), with accurate values $\omega =\gamma=1$.}
    \label{fig:2}
\end{figure}

In the following, we will use the method for accurate force sensing of a levitated nanoparticle. Consider the center-of-mass motion of the nanoparticle along the $x$-axis, with its mass $m$ and frequency $\omega_0$, with the bare Hamiltonian given by
\begin{align}
\hat{H}_0 = \frac{\hat{P}^2}{2m} + \frac{1}{2} m \omega_0^2 \hat{X}^2 = \frac{\hbar \omega_0}{4}(\hat{p}^2 + \hat{x}^2),
\end{align}
where $\hat{X} = x_0 \hat{x}$ and $\hat{P} = p_0 \hat{p}$, with $[\hat{X}, \hat{P}]=i\hbar$ are the position and momentum operators, respectively. $\hat{x}$ and $\hat{p}$ are dimensionless operators, expressed in terms of zero point position $x_0$ and zero-point momentum $p_0 = \hbar/(2x_0)$. Following earlier studies~\cite{Weiss2021_force_sensing, Milburn1999_ion_trap_decoherences}, we consider two types of decoherence channels: (a) statistical displacement of the center of mass of the levitated particle due to fluctuations in the levitation potential, caused by external field fluctuations; and (b) fluctuations in the spring constant resulting from fluctuations in the alternating current component of the applied potential to the trap potential~\cite{ion_trap_decoherences}. This effectively introduces displacement and frequency noise channels, namely ${\gamma_1}\mathcal{D}[\hat{x}]$ and ${\gamma_2}\mathcal{D}[\hat{x}^2]$, where $\gamma_1$ and $\gamma_2$ are the corresponding decoherence rates. The Hamiltonian in the presence of a static force ${F}$ is given by
\begin{equation}
    \hat{H} = \hat{H}_0  + {F}\hat{X} = \hat{H}_0  + \hbar \omega_0 {f}\hat{x},
    \label{eq:qho_f}
\end{equation}
where $f = F x_0/(\hbar \omega_0)$ is the dimensionless static force to be detected. With this, the SME is given by
\begin{align}
\nonumber
{d\rho_c(t)} &=  - i [\hat{H}, ~\rho_c(t)] dt + \left[\gamma_1 \mathcal{D}[\hat{x}] + \gamma_2 \mathcal{D}[\hat{x}^2] \right] \rho_c(t) dt\\
&+ \kappa \mathcal{D}[\hat{a}] \rho_c(t) dt+ \sqrt{\kappa \eta} \mathcal{H}[\hat{a}] \rho_c(t) d\mathcal{W}(t),
\label{eq:SME_nano}
\end{align}
where $\hat{a}$ is the measurement operator in the homodyne measurement scheme, that results in the position $\hat{x} = (\hat{a} + \hat{a}^\dagger)/\sqrt{2}$ as the output signal of the measurement. Nevertheless, the measurement current will be masked by the measurement noise that yields only the noisy measurement output $\mathcal{I}(t) = \langle \hat x (t)\rangle_c + {(4\kappa \eta)}^{-1/2}  d\mathcal{W}/dt$. 
Alternative measurement operators ${\hat A}$, which may be more advantageous for the specific experimental configuration used, are possible. In such a context, too, the protocol remains equally effective.

\begin{figure}[t]
    \centering
    \includegraphics[width=1.0\linewidth]{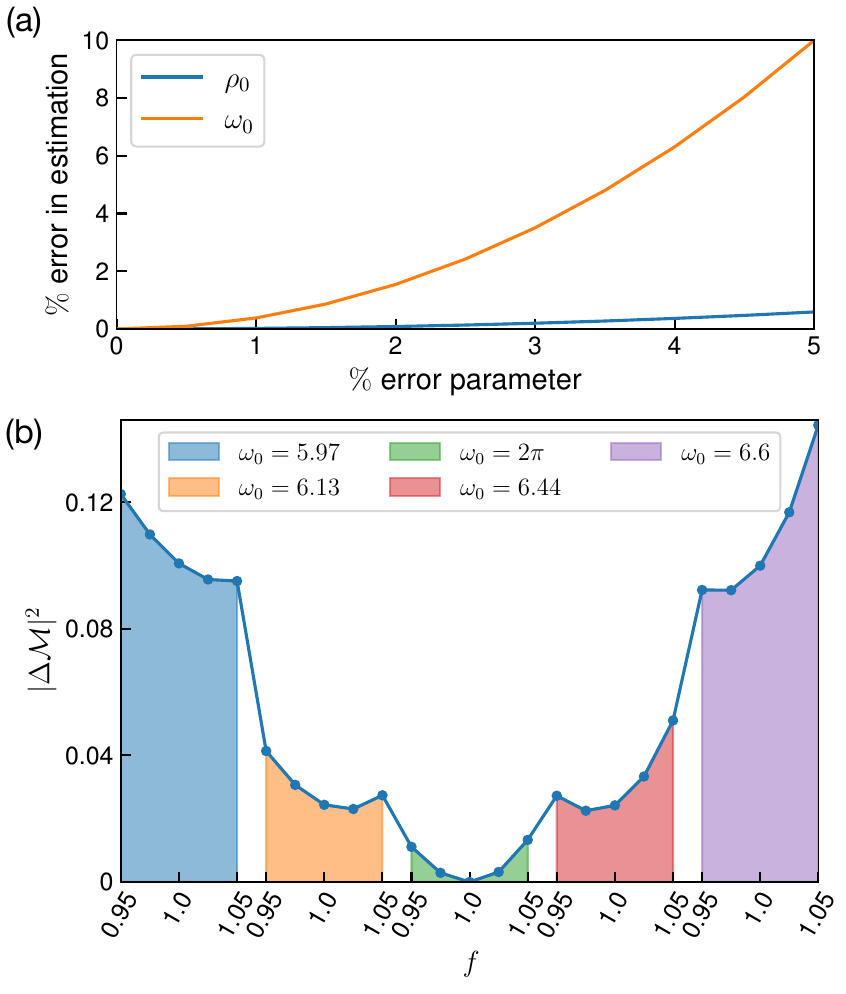}
    \caption{(a) Estimation error (in percentage) due to inaccuracy in parameter (in percentage). We consider the inaccuracy as we deviate from the exact initial state $\rho_0$ and the exact frequency $\omega_0$. $\rho_0$ are considered coherent states defined by the complex amplitude $\alpha$. Although the estimation error for the accurate force $f$ is within 0.5\% upto 5\% of error $\rho_0$ (defined by the error in $\alpha$), it is about 10\% for 5\% change in the value of $\omega_0$. (b) Simultaneous estimation of $\omega_0$ and $f$. Each shaded region defines a value of $\omega_0$ for which $f$ is varied. The accurate pair of values, that is, $(\omega_0, f) = (2\pi, 1.0)$, yields the minimum loss. }
    \label{fig:3}
\end{figure}
Now, as per the procedure, the first step is to collect a homodyne trajectory $\{\mathcal{I}_{\rm expt}(t)\}$ starting from a known initial state $\rho_0$. This trajectory will act as our reference dataset $D$, which in our case is generated by solving the SME given by Eq.~\ref{eq:SME_nano}. The simulator is constructed for this system based on the known system model: $\omega_0$, $\gamma_{1}$, $\gamma_{2}$, $\kappa$ and $\eta$. If they are not known a priori, the same procedure can be applied; for example, we can estimate $\omega_0$ and $\gamma$ beforehand [see Fig.~\ref{fig:2}(e)]. Therefore, the unknown part of the Hamiltonian is $\Delta H = \hat{H} - \hat{H}_0$, which embodies the force to be detected. The next step is to solve the problem as a post-processing step, using Eq.~\ref{eq:mbe_scheme_main_eq} and Eq.~\ref{eq:loss}. The minimum of the loss should ideally be zero, which would correspond to the accurate parameter $f\sim f_0$ to be estimated. 

The results of the force estimation are shown in Fig.~\ref{fig:2}. A typical experimental trajectory is shown in Fig.~\ref{fig:2}(b), which acts as the reference data set for post-processing. The calculated loss is shown as a function of $f$ in Fig.~\ref{fig:2}(c). The minimum of loss [$|\Delta \mathcal{M}|^2 = 0$] makes each point of the experimental trajectory exactly reproduced, as shown in Fig.~\ref{fig:2}(b). If we analyze the conditional means $\langle \hat{x}(t)\rangle$, they are also perfectly converged (shown in bold lines). To quantify the sensitivity of the estimation with respect to the parameter to be estimated, we have calculated the derivative of the force with respect to the loss,  $df/|\Delta \mathcal{M}|^2$, shown in Fig.~\ref{fig:2}(d). 
Fig.~\ref{fig:2}(e) provides an example of the QPE scheme, demonstrating the process of determining the frequency $\omega$ [depicted in blue] of a quantum harmonic oscillator Hamiltonian, $\hat H = \hbar\omega \hat a^{\dagger} \hat a$, where $\hat{a}(\hat{a}^\dagger)$ represents the annihilation (creation) operator of the oscillator. Furthermore, if the system possesses an unknown damping at a rate of $\gamma$, the same procedure can be utilized to estimate it [illustrated in orange]. To simplify matters, we rescale the parameters $\lambda = (\omega, \gamma)$ such that $\lambda_{\rm accurate}=1$. The inset shows a zoomed-in view around $\lambda=1$ demonstrating the fact that the loss $\Delta \mathcal M \to 0$ as $\omega \to \omega_{\rm accurate}$ and $\gamma \to \gamma_{\rm accurate}$.

We show in Fig.~\ref{fig:3}(a), the estimation error resulting from the imprecision in the initial parameter selections of the simulator relative to the experimental parameters. Here, we consider the inaccuracy as we deviate from the exact initial state $\rho_0$ and the exact frequency $\omega_0$. For this exercise, we considered $\rho_0$ as coherent states defined by the complex amplitude $\alpha$, and consider the errors as $\alpha$ deviates from the true value $\alpha_0$.  Although the estimation error for the accurate force $f$ is within 0.5\% up to 5\% of the error $\rho_0$ (defined by the error in $\alpha$), it is about 10\% for the change of 5\% in the value of $\omega_0$. This is rather expected, since $\omega_0$ appears in both terms in the Hamiltonian in Eq.~\ref{eq:qho_f}, and therefore it is expected to be estimated very well in advance following the method [as in Fig.~\ref{fig:2}(e)]. In Fig.~\ref{fig:3}(b), we show the case where $\omega_0$ and $f$ are estimated simultaneously. Each shaded region defines a value of $\omega_0$ for which $f$ is varied. The accurate pair of values, that is, $(\omega_0, f) = (2\pi, 1.0)$, yields the minimum loss. 

\begin{figure}[t]
    \centering
    \includegraphics[width=1.0\linewidth]{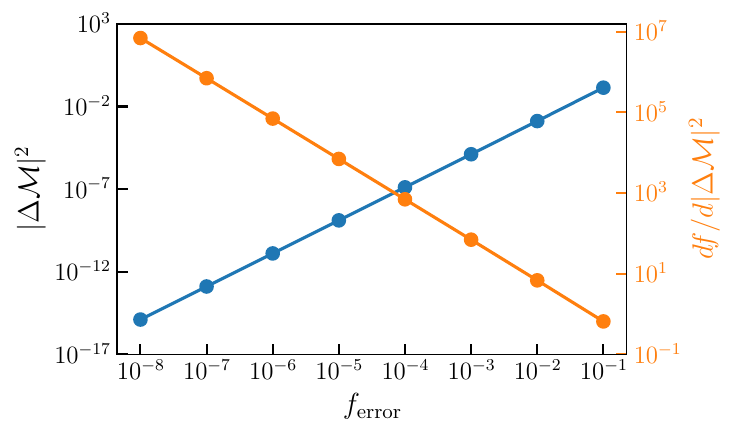}
    \caption{The loss as a function of the error in estimation of force, $f_{\rm error} = f - f_{0}$. $f_{\rm error}=0$ is not shown for which $|\Delta \mathcal{M}|^2 = 0$. On the twin-axis, the corresponding derivative $df/d |\Delta \mathcal{M}|^2$ is shown. The $y$-axes are in logarithmic scale.}
    \label{fig:4}
\end{figure}

In Fig.~\ref{fig:4}, we quantify the accuracy in estimating $f$. For this, we plot the loss $|\Delta \mathcal{M}|^2$ [Eq.~\ref{eq:loss}] as a function of the error in predicting $f$, i.e. $f_{\rm error} = f - f_{0}$. We see that the loss decreases linearly with the decrease in $f_{\rm error}$ with $|\Delta \mathcal{M}|^2 \to 0$ as $f_{\rm error} \to 0$. Therefore, the method outlined above is perfectly accurate, provided that we have perfect accuracy in the model parameters. 

The Quantum Cram\'{e}r-Rao Bound (QCRB) is the fundamental bound for parameter estimation. Here, we show that our estimation of a constant force obeys the QCRB, even though the accuracy is perfect. The QCRB can be expressed as a spectral uncertainty principle in our case \cite{Caves2011Fundamental}:
\begin{align}
    S_{f_\mathrm{error}}(\omega) \geq \left(\frac{4}{\hbar^2}|\chi_\mathrm{m}(\omega)|^2 S_{f_\mathrm{ba}}(\omega)+\frac{1}{S_f(\omega)}\right)^{-1},
\end{align}
where $S_{f_\mathrm{error}}(\omega)$ is the power spectral density of the estimation error of the force $f$, $\chi_\mathrm{m}(\omega)$ the mechanical susceptibility of the levitated particle, $S_{f_\mathrm{ba}}(\omega)$ the spectrum of the backaction force, and $S_f(\omega)$ the prior spectrum of the force $f$. As $f$ is a static force, $S_f(\omega)=f^2\delta(0)$. One can also write the bound in terms of variance,
\begin{align}
    V_{f_\mathrm{error}} \geq \int_{-\infty}^{\infty} \frac{d\omega}{2\pi}\left(\frac{4}{\hbar^2}|\chi_\mathrm{m}(\omega)|^2 S_{f_\mathrm{ba}}(\omega)+\frac{1}{S_f(\omega)}\right)^{-1},
\end{align}
where $V_{f_\mathrm{error}}$ is the variance of $f_\mathrm{error}$.

For given measurement noise, the estimation accuracy is limited by a weaker bound. The measurement current in the frequency domain $\mathcal{I}_\mathrm{expt}(\omega)$ can be written in displacement unit as $\hat{y}(\omega)=\chi_\mathrm{m}(\omega)(f(\omega)+\hat{z}(\omega))$, where $\hat{z}(\omega)$ contains all force noises, including quantum backaction noise and imprecision noise expressed in force unit. Using the smoothing technique, one can get the minimum achievable error for a given force noise spectrum $S_z(\omega)$\cite{Tsang2009Time,Caves2011Fundamental}:
\begin{align}\label{eq:Vf}
    V_{f_\mathrm{error}}^\mathrm{min} = \int_{-\infty}^{\infty} \frac{d\omega}{2\pi}\left(\frac{1}{S_z(\omega)}+\frac{1}{S_f(\omega)}\right)^{-1},
\end{align}
which is known to obey the QCRB. 

Our estimation method is based on the vanishing loss function in Eq.~\ref{eq:loss}, which can be written as 
\begin{align}
    {\Delta \mathcal{M}^2(\omega)} = |\hat{y}_{\rm expt}(\omega) - \hat{y}_{\rm QPE}(\omega)|^2
    = |f_\mathrm{error}(\omega) + \hat{z}_\mathrm{error}(\omega)|^2,
\end{align}
where $f_\mathrm{error} = f_{\rm QPE}-f_0$, similarly $\hat{z}_\mathrm{error} = \hat{z}_{\rm QPE}-\hat{z}_{\rm expt}$. $\Delta\mathcal{M}^2(\omega)=0$ infers perfectly anticorrelated $f_\mathrm{error}$ and $\hat{z}_\mathrm{error}$, that is, $f_\mathrm{error}=-\hat{z}_\mathrm{error}$. Then the spectrum of the estimation error $S_{f_\mathrm{error}}=S_{\hat{z}_\mathrm{error}}$. As $z_\mathrm{expt}$ and $z_\mathrm{QPE}$ are in the experiment and simulation, respectively, they are not correlated. Therefore, $S_{f_\mathrm{error}}(\omega)=S_{\hat{z}_\mathrm{error}}(\omega)=2S_z(\omega)$. Notice that the integrand of Eq.~\ref{eq:Vf} $\left(1/S_z(\omega)+1/S_f(\omega)\right)^{-1}<S_z(\omega)<S_{f_\mathrm{error}}(\omega)$, our method does not violate the QCRB. This also implies the following expression of our estimation error variance:
\begin{align}
    V_{f_\mathrm{error}} = \int_{-\infty}^{\infty} \frac{d\omega}{2\pi}2S_z(\omega)B(\omega),
\end{align}
where $B(\omega)$ defines the effective bandwidth of the force to be estimated. This effective bandwidth is encoded in the $S_f(\omega)$ in the expression of QCRB. Outside the bandwidth, $S_f(\omega)=0$, and the integrand vanishes. 
Intuitively, applying $B(\omega)$ reduces the bandwidth of the integrated noise, which is favorable for parameter estimation. 

Obeying the QCRB still allows conceptually perfect accuracy of estimation in the sense of variance, due to the zero-bandwidth of the static force to be estimated, i.e. $B(\omega)=0$. In the vicinity of 0, as $S_z(0)$ is finite, the integration of $S_{f_\mathrm{error}}=2S_z(\omega)$ around 0 vanishes. Therefore, $V_{f_\mathrm{error}}=0$, i.e.~the accuracy of the estimate is perfect.
In practice, the experimental spectrum does not have infinite resolution, and the integration bandwidth is limited by $1/\tau$, where $\tau$ is the length of the trajectory in time.
This gives a finite estimation accuracy, but scales favorably with integration time.

In conclusion, accurate estimation of parameters in experiments in quantum systems is important from the perspective of a host of applications. In this work, we have introduced a method that can be useful for estimating and sensing parameters from a single quantum trajectory from an experiment that undergoes continuous measurement. As a useful and intuitive demonstration, we have shown its usefulness for the application of parameter estimation and force sensing of a levitated nanoparticle, a field that has become of immense importance in recent years because of the possibilities of realizing them in laboratory experiments on different platforms, such as optical/ion traps and magnetic levitation.  However, the same method can also be applied for estimating parameters in other systems and quantum platforms.

% \section*{Acknowledgements} 


% \bibliographystyle{unsrt}
% \bibliography{bibfile}

\begin{thebibliography}{32}
\bibitem [{ion()}]{ion_trap_decoherences}%
  \BibitemOpen
  \href@noop {} {}\bibinfo {note} {Ion traps utilize electromagnetic fields to confine ions. External electric fields can fluctuate due to various sources such as ambient electromagnetic interference, imperfections in the trapping apparatus, and constraints in the control electronics. These fluctuations induce a time-dependent force on the ion, leading to linear decoherence in position, $x$. Consequently, the mean energy increases linearly with time. Conversely, the trapping potential in ion traps can vary due to alterations in the applied voltages, both AC and DC. These variations affect the spring constant of the potential, resulting in quadratic decoherence in position, $x^2$. This leads to an exponential increase in energy.}\BibitemShut {Stop}%
\end{thebibliography}

\end{document}